\documentclass[aps,twocolumn,prl,tightenlines,floatfix,showpacs]{revtex4-1}

\usepackage[dvips]{graphicx}
\usepackage[english]{babel}
\usepackage{amsmath}
\usepackage{amssymb}
\usepackage{times}

\begin{document}

\title{Dynamics of a quantum quench in an ultra-cold atomic BCS superfluid}
\author{Chih-Chun Chien and Bogdan Damski}
\affiliation{Theoretical Division, MS B213, Los Alamos National Laboratory, Los Alamos, NM, 87545, U.S.A.}

\begin{abstract}
We study dynamics of  an  ultra-cold atomic BCS superfluid 
driven towards the BCS superfluid-Fermi liquid 
quantum critical point  by a gradual  decrease of the 
pairing interaction. We analyze how the BCS superfluid falls 
out of equilibrium and show that the non-equilibrium gap and Cooper pair size
reflect critical properties of the transition.  
We observe three stages of evolution: 
adiabatic where the Cooper pair size is inversely proportional to the equilibrium gap, 
weakly non-equilibrium  where it is inversely proportional to 
the non-equilibrium gap, 
and strongly non-equilibrium where it decouples from 
both equilibrium and non-equilibrium  gap. 
These phenomena should stimulate future experimental characterization of non-equilibrium ultra-cold atomic 
BCS superfluids.
\end{abstract}


\maketitle

Ultra-cold atoms provide quantum simulators for emulating challenging 
theoretical models. This offers invaluable opportunities for 
the study of many-body systems which used to be experimentally
inaccessible by conventional condensed matter setups \cite{LewensteinReview}. 
The ability to make the parameters of the 
emulated models time-dependent encourages exciting studies of 
non-equilibrium dynamics of many-body quantum systems \cite{DziarmagaReview}.
This opens up fresh prospects for interdisciplinary research 
of dynamics of phase transitions, where universality 
of the critical behavior links dynamics of distinct  systems 
quenched across a critical point \cite{Zurek}. Examples include superfluids \cite{superfl}, superconductors \cite{Maniv2003}, 
spin chains \cite{DziarmagaReview}, cosmological models \cite{Kibble},  and others \cite{Zurek}.

In conventional superconductors, observations of the dynamics of superconductivity would 
require a real-time change of either dopant concentration or 
lattice structure, injection of quasi-particles, or exposition to microwave or light \cite{Tinkham,JJChang}, which is very difficult. 
The dynamics of BCS wavefunction is also of interest to nuclear physicists \cite{Koonin}.
Such dynamics can be realized in cold atoms by adjusting  the inter-atomic 
interactions, which is routinely
done  in current experiments. Moreover, the dynamics of a BCS superfluid can be mapped to a central-spin problem \cite{YuzbashyanJPhys}. 
Our goal is to study the non-equilibrium dynamics of an $s$-wave BCS superfluid
resulting from a ramp of attractive pairing interactions.

Previous work on the quench dynamics of a BCS superfluid focused 
on the non-equilibrium gap following 
an instantaneous  change of the pairing interaction 
\cite{LevitovT0,Yuzbashyan,Burnett}. 
Here we explore finite-rate quench dynamics which, unlike an 
instantaneous one, allows for an observation of all stages of 
the evolution ranging from  adiabatic to strongly non-equilibrium ones.
This can be done in a controllable way by changing
the quench rate, which also allows for observations 
of universal critical properties of the phase transition 
through the scaling relations of out-of-equilibrium 
observables. 
These experimentally accessible features should make a substantial progress in the understanding of non-equilibrium 
physics of BCS superfluids.  A linear ramp of the external magnetic field around the unitary point is predicted to drive the pairing gap out of equilibrium \cite{Yi}.

Several fundamentally important questions will  be addressed:  
What is the non-equilibrium gap and Cooper pair size during the ramp of the pairing interactions and how are the two quantities related? 
How do those non-equilibrium quantities 
reflect the critical properties characterizing the 
BCS superfluid-Fermi liquid (BCS-FL)
phase transition?

We consider a three-dimensional homogeneous two-component ($\sigma=\uparrow$, $\downarrow$) unpolarized Fermi gas.
Although in experiments there is a trap potential, the center of the Fermi gas can be considered as locally homogeneous and the dynamics we investigate should be observable. For weakly attractive or repulsive  interactions the ground state is a BCS superfluid \cite{PethickBEC} or a normal Fermi liquid \cite{CCCFL}, respectively. The Hamiltonian with a time-dependent BCS-type pairing interaction is  
\begin{equation}
\hat H = \sum_{k\sigma} \epsilon_k\hat c^\dag_{k\sigma}\hat c_{k\sigma}
         - \lambda(t)\sum_{kl} \hat c^\dag_{k\uparrow}
		                    \hat c^\dag_{-k\downarrow}
							\hat c_{-l\downarrow}
							\hat c_{l\uparrow}.
\end{equation}
Here $\epsilon_k=\hbar^{2}k^2/2m$ and 
$\lambda(t)$ is the time-dependent coupling constant from interactions
between atoms in different hyperfine states,  
which is regularized as 
$\lambda^{-1}=-m/4\pi \hbar^2 a+\sum_{k<{\cal K}}(1/2\epsilon_k)$ \cite{Leggett}. 
$a$ is the two-body $s$-wave scattering length and ${\cal K}\gg k_F$ 
is a momentum cutoff \cite{numerics}. 
We choose 
as our units 
the Fermi momentum and energy $k_F$ and $E_F$ of a non-interacting Fermi gas with the same density.
Experimentally $a$ is tuned by an external magnetic field $B$ via  
$a=a_{bg}[1-\delta B/(B-B_0)]$,
where $a_{bg}$ is the background scattering length, while 
$\delta B$ and $B_0$ are the width and position of the Feshbach resonance 
\cite{PethickBEC}. The $T=0$ BCS-FL quantum critical point corresponds to the point where the system becomes a non-interacting (free) Fermi gas.
In BCS-Leggett theory the critical point is at $a=0$ and it can be located numerically for a finite-range potential \cite{finiterange}. This critical point for $^{40}$K atoms is located at $209.9G$ \cite{Jin_a,FermionMott}.

The excitation gap on the BCS side ($a<0$) closes exponentially  near 
the critical point. It 
is well approximated for $k_Fa>-1$ by \cite{Leggett}
\begin{equation}
\Delta^{\rm eq}\approx8E_Fe^{-\pi/2k_F|a|}/e^2.
\label{gap}
\end{equation}
Importantly, the excitation gap in BCS theory  serves
as the order parameter as well. This is a feature
that distinguishes a BCS superfluid from other systems previously 
studied in the field of dynamics of quantum phase transitions \cite{DziarmagaReview}.

The BCS coherence length (after adjusting an overall constant) behaves 
near the quantum critical point as \cite{Tinkham}
\begin{equation}
k_F\xi^{\rm eq}= E_F/\sqrt{2}\Delta^{\rm eq}.
\label{coh}
\end{equation}
Equations (\ref{gap}) and (\ref{coh}) set the critical exponents 
to $z=1$ and $\nu=\infty$ through the relations 
$\xi^{\rm eq} \sim (\Delta^{\rm eq})^{-1/z}$ and $\xi^{\rm eq}\sim|a|^{-\nu}$.
The infinite $\nu$ exponent makes a BCS superfluid distinct from typical
 systems considered  in the dynamics
of phase transitions where $\nu = {\cal O}(1)$ \cite{DziarmagaReview}.
To account for this singular behavior
we will propose a quench protocol allowing for efficient tests of quench-induced
scaling relations.

To numerically study the effect of driving, we employ the generalized BCS-Leggett wavefunction
\cite{Leggett}
\begin{equation}
|\Psi(t)\rangle = \prod_k(u_k(t) + v_k(t)\hat c^\dag_{k\uparrow}\hat c^\dag_{-k\downarrow})
               |0\rangle,
\label{BCSwave}
\end{equation}
which provides a consistent description of 
BCS-Bose Einstein condensate (BEC)
crossover but does not 
include medium effects \cite{LevitovT0,Yuzbashyan,Burnett,PethickBEC,Leggett,KetterlePsize}, which renormalize the equilibrium gap.
The medium effects can introduce corrections to the dynamics, which may be evaluated non-perturbatively beyond BCS-Leggett theory \cite{PethickBEC}.
Minimizing the quantum action 
$
\int dt (i\hbar\langle\Psi|\partial_t|\Psi\rangle/2+c.c.
                       -\langle\Psi|\hat H|\Psi\rangle)
$
 \cite{details}, we obtain 
the following equations:
\begin{equation}
\hbar\dot d_k = 2ig_k\Delta + 2id_k\epsilon_k, \ \ 
\hbar\dot g_k = id_k\Delta^* - id_k^*\Delta.
\label{equations}
\end{equation}
Here 
$d_k=2u_kv_k^*$, $g_k=|v_k|^2-|u_k|^2$, and the non-equilibrium 
gap function $\Delta(t)=\lambda(t)\sum_k d_k(t)/2$. 
Eqs.~(\ref{equations}) 
are consistent with those in Ref.~\cite{Burnett}. 
Here $\tau_0=\hbar/E_F$ is the unit of time. 
Eqs.~(\ref{BCSwave}) and (\ref{equations}) can be applied to the whole BCS-BEC crossover, 
though here we focus on the quench dynamics close to the $a=0$ critical point.

The quench dynamics  proceeds as follows. 
Initially the system is in its ground state at $k_Fa\ll-1$, and
the pairing interaction is slowly decreased towards
$a=0$ (the quantum critical point) by tuning away from the Feshbach resonance. 
In the beginning  the system evolves  adiabatically. 
At some point, say $a=-\hat a$, the 
gap becomes so small that the system 
will fall out of equilibrium. 
This happens when the reaction time $\hbar/\Delta^{\rm eq}$ of the BCS superfluid 
 becomes comparable to the inverse of the quench rate $\Delta^{\rm eq}/|\dot\Delta^{\rm eq}|$ 
 \cite{quantum_quench}: 
\begin{equation}
\frac{\hbar}{\Delta^{\rm eq}(-\hat a)}
={\cal C} \left.\frac{\Delta^{\rm eq}}{|\dot\Delta^{\rm eq}|}\right|_{a=-\hat a},
\label{timescales}
\end{equation}
where ${\cal C}$ is a constant and the dot stands for $d/dt$. 

Basic understanding of what happens next comes from the 
Kibble-Zurek theory \cite{Kibble,Zurek}, 
which has not been verified for a BCS superfluid yet.
This theory predicts that  a system switches suddenly from adiabatic to diabatic
dynamics:  its state is frozen  near the critical point. 
Thus in the non-equilibrium 
stage of the evolution, the order parameter $\Delta$ and coherence length 
$\xi$ should be given by their values $\hat \Delta$ and $\hat \xi$ at the border between 
adiabatic and diabatic regimes:  
$
\hat\Delta = \Delta^{\rm eq}(-\hat a), \ \ \hat\xi = \xi^{\rm eq}(-\hat a).
$
In reality there is no sudden freeze out, but one can investigate 
{\it if} the following scaling relations hold near the critical point: 
\begin{equation}
\Delta(t) = \hat\Delta f(t/\hat t\,), \ \ \xi(t) = \hat\xi g(t/\hat t\,),
\label{near_crit}
\end{equation}
where $f$ and $g$ are unknown functions,  
$\hat t$ is the time left to reaching the critical point
at the onset of non-equilibrium dynamics, $a(-\hat t) = -\hat a$,
and the critical point is reached at $t=0$.
Once the quench protocol, $a(t)$, 
is specified, quantitative predictions about system dynamics can be made.

\begin{figure}
\includegraphics[width=\columnwidth,clip=true]{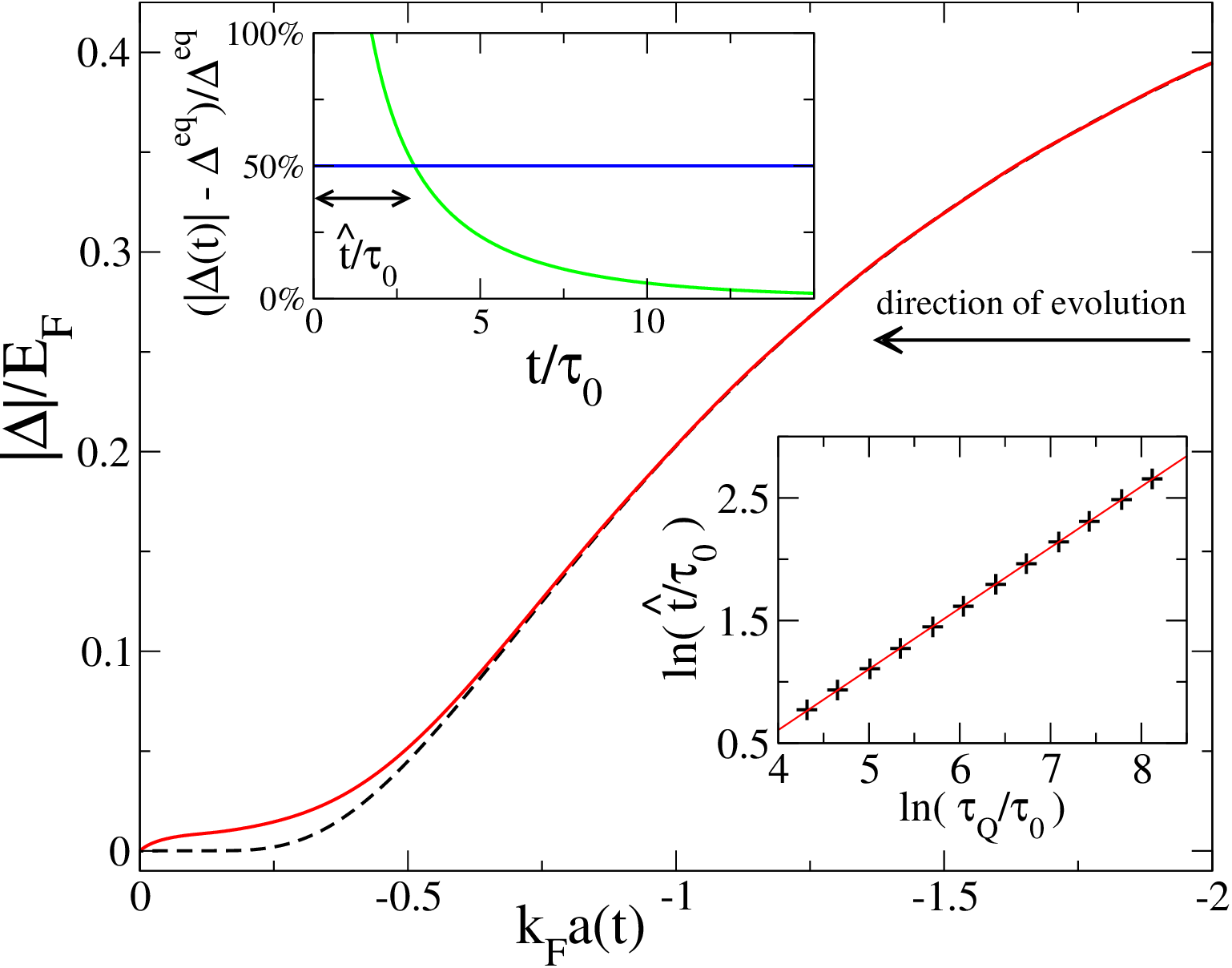}
\caption{(Color online) Non-equilibrium dynamics of the gap.
         Red solid line shows $|\Delta(t)|$ for $\tau_Q=150\tau_0$ while 
		 black dashed line depicts $\Delta^{\rm eq}$. The two results
		 start to deviate near $k_Fa(t)=-0.75$, which signals the onset of
		 non-equilibrium dynamics. 
		 Upper inset: Illustration of 
	      how the threshold of non-equilibrium gap dynamics is 
		 determined. The solid green curve  
		 shows the relative difference between $\Delta^{\rm eq}$
		 and non-equilibrium gap for $\tau_Q=150\tau_0$. 
		 When the relative difference reaches $50\%$,
		 the time left to reach the critical point equals
		 $\hat t$. The power-law scaling of $\hat t$
		 is shown in the lower inset. The linear fit (solid red line) 
		 to numerical data (pluses) gives  
		 $\ln(\hat t/\tau_0)=-1.38\pm0.005 + (0.496\pm0.001)\ln(\tau_Q/\tau_0)$. It is
		 in good agreement with the prediction $\hat t \sim \tau_Q^{0.5}$ (\ref{new_hats}).	 
		 See \cite{numerics} for details. 
}
\label{fig1}
\end{figure}

The linear ramp of the external magnetic field proposed in Ref.~\cite{Yi} results in a non-linear drive of $k_Fa(t)$ and makes it difficult to predict the dynamics. 
 A more typical quench protocol assumes a linear quench \cite{DziarmagaReview}. 
This corresponds here
to having $k_Fa(t) \propto t/\tau_Q$, 
where $\tau_Q$ provides the quench time-scale: the slower the system is driven,
the larger $\tau_Q$ is. Such a quench 
can be realized by 
a non-linear (in time) sweep of $B(t)$ near a Feshbach resonance.
For a system with an exponentially closing gap, however, it introduces
logarithmic corrections (in $\tau_Q$) to the non-equilibrium length scale 
$\hat\xi$ \cite{Polkovnikov2005}, 
which demands very long quench times for verification of scaling relations.
To overcome this problem, we propose the following quench protocol 
\begin{equation}
\Delta^{eq}(a(t))e^2/8E_F = -t/\tau_Q,
\label{new_dt}
\end{equation}
where time goes from $t_i<0$ to $0$ such that $k_Fa(t_i)<-1$, and the 
$e^2/8$ prefactor is for convenience.
This can be inverted for $k_Fa>-1$ using Eq.~(\ref{gap}) 
\begin{equation}
k_Fa(t) = \pi/\ln(t^2/\tau_Q^2).
\label{new_at}
\end{equation}
Such a ramp may be induced by 
$B(t)=B_{0}+\delta B/[1-\pi/k_{F}a_{bg}\ln(t^2/\tau_Q^2)]$. To avoid inessential complications with the exact 
inversion of Eq.~(\ref{new_dt}), 
we use the protocol shown in Eq.~(\ref{new_at})
through the whole evolution and choose the quench times so slow
that the system is brought to $k_Fa=-1$ nearly adiabatically.
Therefore, significant excitations in our calculations only happen at 
$k_Fa>-1$, where Eqs.~(\ref{new_dt}) 
and (\ref{new_at}) become consistent.
From Eqs.~(\ref{timescales}) and (\ref{new_dt}) one gets
\begin{eqnarray}
&& k_F\hat a = \pi/\ln(8{\cal C}\tau_Q/\tau_0e^2), \ \ 
\hat\Delta\propto\sqrt{\tau_0/\tau_Q}E_F, \nonumber \\ 
&& \hat t\propto\sqrt{\tau_Q\tau_0}, \ \   
k_F\hat\xi\propto \sqrt{ \tau_Q/\tau_0}.
\label{new_hats}
\end{eqnarray}
Eqs. (\ref{new_hats})  complement the scaling relations shown in Eqs.~(\ref{near_crit}).
A simple power-law dependence of $\hat\Delta$, $\hat t$ and $\hat \xi$ 
on the quench time scale  
 will serve as a verification of our predictions.

Typical dynamics of the gap function is illustrated 
in Fig. \ref{fig1}. Away from the critical point the gap is large
enough to enforce adiabatic evolution in which $\Delta$
obtained from  Eq.~(\ref{equations})  matches $\Delta^{\rm eq}$.
As the gap becomes small enough, 
the relative deviation, 
$(|\Delta|-\Delta^{\rm eq})/\Delta^{\rm eq}$, grows fast (see the upper inset 
of Fig. \ref{fig1}). 
From the simulations, this happens when the time 
left to reach the critical point, $\hat t$, behaves as 
$\tau_Q^{0.5}$, which agrees with the prediction of Eq.~(\ref{new_hats}). 
Moreover, the difference $|\Delta|-\Delta^{\rm eq}$ is controlled
by the quench rate and increases as $\tau_Q$ decreases. 
Finally, we have checked that the scaling relation 
for $\Delta$ near the critical point, Eq.~(\ref{near_crit}), holds.
Very close to the $a=0$ critical point both $\Delta$ and 
$\Delta^{\rm eq}$ approach zero. 
This reflects the vanishing coupling constant $\lambda(t)$  
and does not imply that the system has reached an equilibrium state.

Equilibrium gap can be inferred from radio-frequency (RF) spectroscopy.
Modification of this technique to non-equilibrium systems has been proposed \cite{Dzero2007}. 
Therefore we expect that the non-equilibrium 
dynamics of the gap function described above will be measured experimentally.

\begin{figure}
\includegraphics[width=\columnwidth,clip=true]{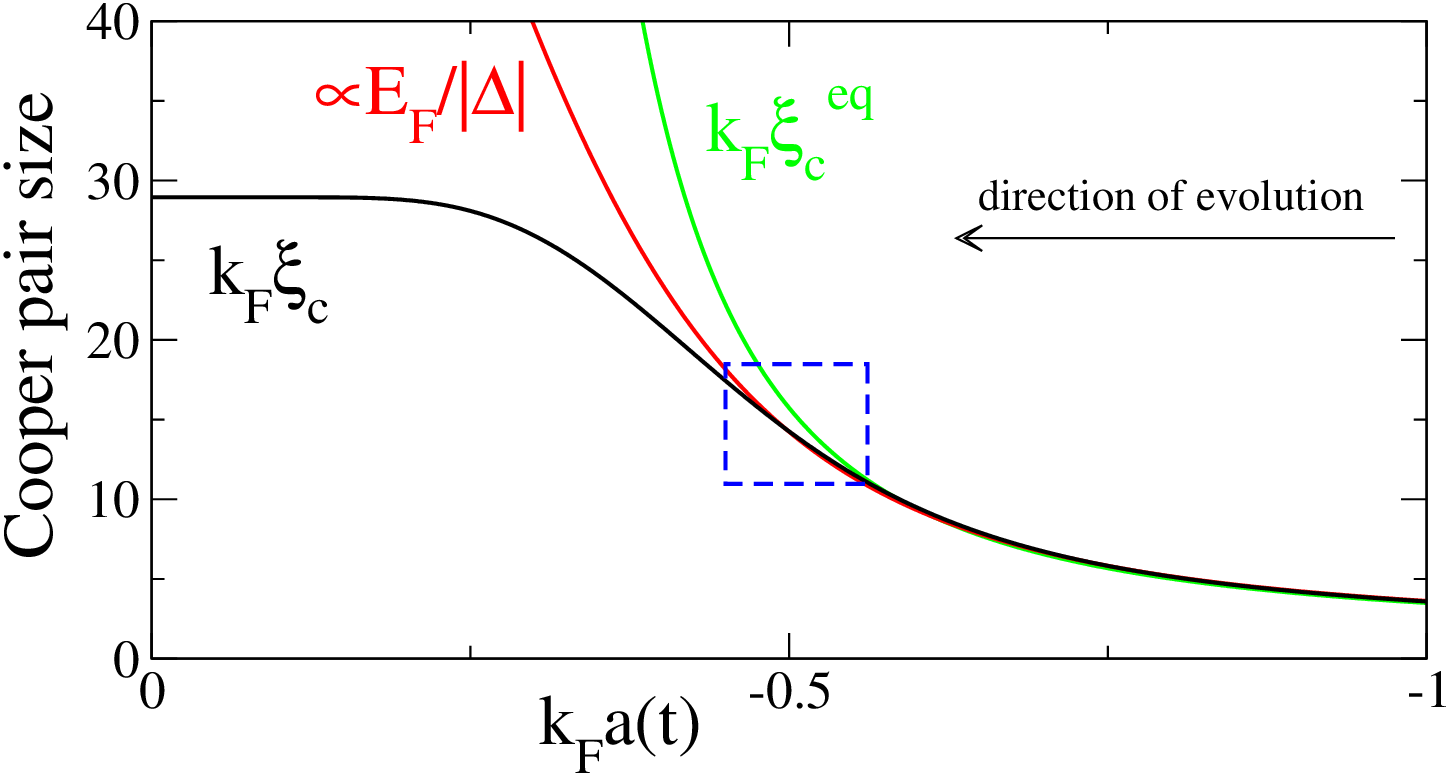}
\caption{
         (Color online) Dynamics of the Cooper pair size in 
         the non-equilibrium quench with $\tau_Q=150\tau_0$.  
		 The curves, from left to right, 
		 correspond to the 
		 non-equilibrium Cooper pair size $k_F\xi_c$ (black),  
		 the inverse of the nonequilibrium gap, $\propto E_F/|\Delta|$ (red), and 
		 the instantaneous equilibrium 
		 pair size $k_F\xi_c^{\rm eq}$ (green). The blue box highlights the region 
		 where $\xi_c\neq\xi^{\rm eq}_c$  
		 but $\xi_c\propto E_F/|\Delta|$. The proportionality factor
		 for the middle (red) curve  was chosen to match the static result 
		 at $k_Fa=-4$ (not shown on the plot).
		 See \cite{numerics} for details. 
}
\label{fig2}
\end{figure}

Next we study the Cooper pair size, which is 
defined as  
$$
\xi_c = \sqrt{\langle\phi| r^2|\phi\rangle/\langle\phi|\phi \rangle}=
\sqrt{\langle\phi|-\nabla_{k}^2|\phi\rangle/\langle\phi|\phi \rangle},
$$
where $|\phi\rangle$ stands for the Cooper pair wave-function:
$\langle k|\phi\rangle = u_kv_k^*$ \cite{PethickBEC,regularization}.
In equilibrium the Cooper pair size $\xi^{\rm eq}_c$ is
the same as the coherence length: $\xi_c^{\rm eq}=\xi^{\rm eq}$. 
In a non-equilibrium situation it is interesting to address
if the Cooper pair size encodes a non-equilibrium 
length scale and remains finite despite the fact 
that $\xi_c^{\rm eq}$ diverges near the critical point.

Typical evolution of $\xi_c$ is depicted in Fig. \ref{fig2}.
It consists of three consecutive stages.
In the first one, the system is away from the critical point and it 
evolves adiabatically: $\xi_c\approx\xi^{\rm eq}_c$. 
In the second one, the system enters non-equilibrium 
dynamics and so $\xi_c\neq\xi^{\rm eq}_c$. Interestingly, we find 
$k_F\xi_c\sim E_F/|\Delta|$ there. Thus, the relation between the 
non-equilibrium Cooper pair size and $|\Delta|$
is the same as that between $\xi_{c}^{\rm eq}$ and $\Delta^{\rm eq}$.
Importantly, it implies that the measurement of the non-equilibrium 
gap proposed in \cite{Dzero2007} can be implemented to estimate the non-equilibrium Cooper pair size.

\begin{figure}
\includegraphics[width=\columnwidth,clip=true]{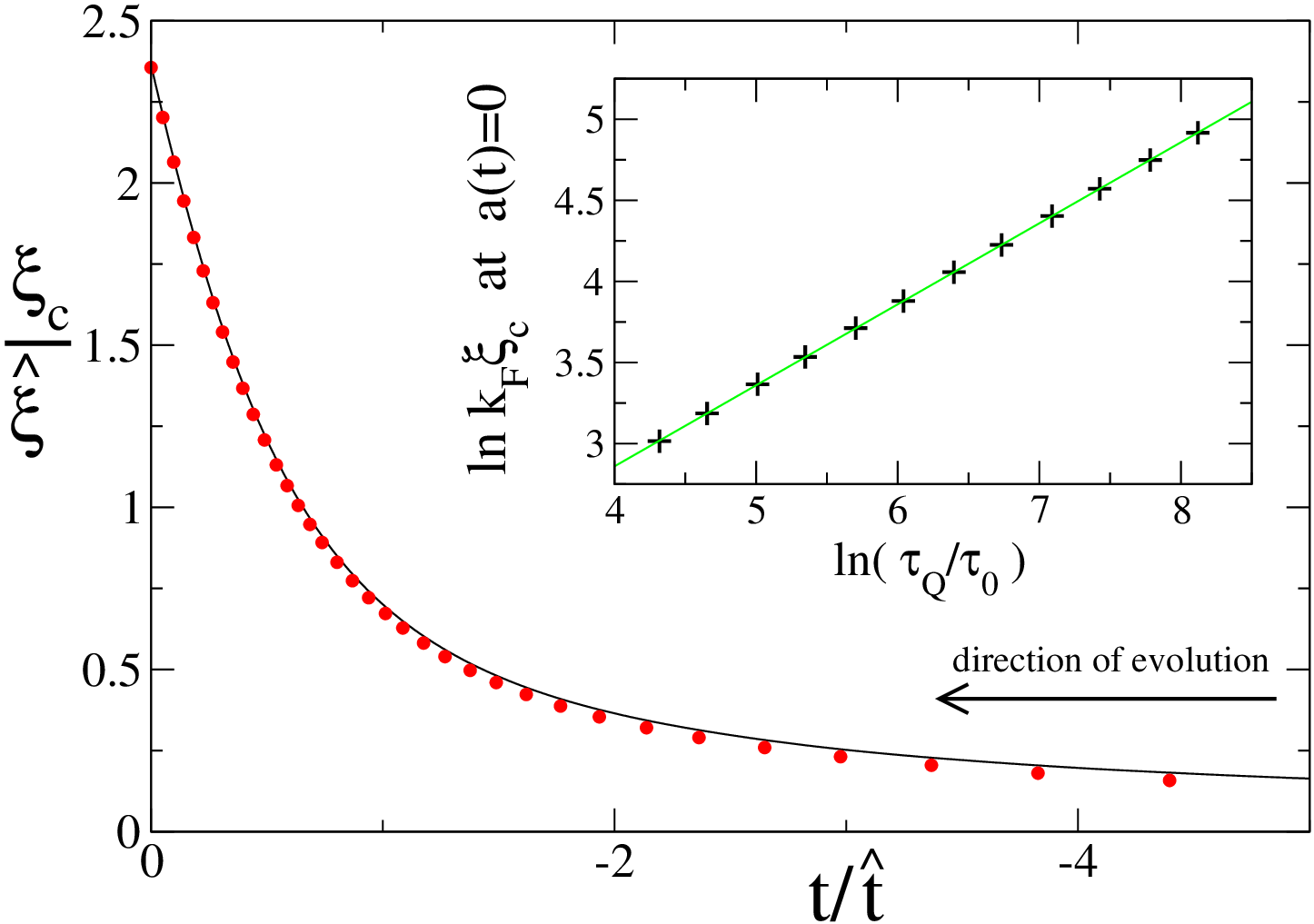}
\caption{
         (Color online) Scaling of the non-equilibrium 
		 Cooper pair size, $\xi_c$, 
		 for $\tau_Q=150\tau_0$ (black line) and for 
		 $\tau_Q=3360\tau_0$ (red dots). Both curves collapse into a universal one  
		 near the quantum critical point, which confirms that the  
		 quench-imprinted non-equilibrium length scale 
		 $\hat\xi$ has not been ``erased'' 
		 until the end of the evolution ($t\to0$). 
		 Rescaling was done with $k_F\hat\xi=\sqrt{\tau_Q/\tau_0}$ and 
		 $\hat t = \sqrt{\tau_Q\tau_0}$. Inset: $\xi_c$ 
		 at the critical point, $a(t=0)=0$. Numerical data are shown as pluses. 
		 They are fitted by  
		 $\ln(k_F\xi_c(a=0))= 0.861\pm0.002 +	 
		 (0.4995\pm0.0003)\ln(\tau_Q/\tau_0)$ (green line). 
		 The expected scaling is $\xi_c(a=0)\sim\tau_Q^{0.5}$.
		 See \cite{numerics} for details. 
		 }
\label{fig3}
\end{figure}

In the third stage both 
$k_F\xi^{\rm eq}_c$ and $E_F/|\Delta|$ diverge, but the non-equilibrium pair size
stays finite. Now $\xi_c$ decouples from the gap function and depends solely
on the non-equilibrium  scales $\hat\xi$ and $\hat t$ through the relation shown in 
Eq.~(\ref{near_crit}), which is depicted in Fig. \ref{fig3}. The inset  of Fig. \ref{fig3}
quantitatively
confirms that the non-equilibrium length scale $\hat\xi$ persists in the system 
till the end of evolution and scales 
as $\tau_Q^{0.5}$, which agrees with the prediction of Eqs.~(\ref{new_hats}).
Surprisingly, the existence of the two stages of non-equilibrium dynamics
is not predicted by the Kibble-Zurek theory.

How to measure the non-equilibrium Cooper 
pair size is an important open question. 
There are experiments  ``measuring'' Cooper pair size by 
inverting the gap obtained from RF spectroscopy \cite{KetterlePsize}.
This will not work close to the critical point (third stage of evolution on Fig.~\ref{fig2})
where $\xi_c$ and $\Delta$ are decoupled. 
One possibility is to generalize to a non-equilibrium 
setting a scheme from \cite{Timmermans2007},
where $\xi_c$ is inferred from the damping rate of the collective mode of a coexisting 
BEC.

We have showed that close to the BCS-FL quantum phase transition at $T=0$, BCS superfluids exhibit multi-stage dynamics with 
scaling behavior
consistent with Kibble-Zurek theory. 
At finite temperature two additional  
effects show up. First, there is a classical (thermal) phase transition 
from a BCS superfluid to a normal Fermi gas at 
$k_BT_c\approx0.57\Delta^{\rm eq}(T=0)$ \cite{Tinkham}, where $k_B$ is the Boltzmann constant. 
Second, quasi-particles (QPs) from broken Cooper pairs exist at $T>0$.
Their relaxation time 
is estimated to be 
$\tau_{\rm qp} \approx \hbar E_F/(\Delta_{\rm GL}^{\rm eq})^2$ \cite{LevitovT0},
where the Ginzburg-Landau gap function $\Delta^{\rm eq}_{\rm GL}$ 
corresponds to the effective gap in the presence of both QPs 
and Cooper pairs. Below we assume that the temperature changes
insignificantly during the slow driving and estimate the influence of these 
finite temperature effects.

If we start the quench dynamics from an equilibrium state 
at $T\ll T_c(a)$ and $k_Fa\ll0$, $\Delta^{\rm eq}_{\rm GL}\approx\Delta^{\rm eq}$ 
initially and the quasi-particle relaxation time $\tau_{\rm qp}$ will
be short compared to the inverse of the quench rate $\Delta^{\rm eq}_{\rm GL}/|\dot\Delta^{\rm eq}_{\rm GL}|$. 
Thus both the superfluid and QPs follow the 
ramp adiabatically. Comparing the two time scales as we did in Eq.~(\ref{timescales}), QPs will stop following the 
quench dynamics of Eq.~(\ref{new_at}) when  the time left to reach the quantum 
critical point is $t^{*}\sim\sqrt[3]{\tau_Q^2\tau_0}$.
Comparing this to the onset of the non-equilibrium dynamics of the condensate, $\hat t\sim \sqrt{\tau_Q\tau_0}$, 
one can see that for $\tau_Q/\tau_0\gg1$ studied in this paper 
the QPs will fall out of equilibrium \textit{before} the 
superfluid does ($t^{*}>\hat t$). Here we 
assume $\Delta^{\rm eq}_{\rm GL}\approx\Delta^{\rm eq}$, which should be reasonable  if  
the system is away from the classical critical point at time  
$t^{*}$. This requires that  
$T\ll T_c(t^{*})\sim\sqrt[3]{\tau_0/\tau_Q}T_F$, where $T_F=E_F/k_B$. 
After QPs fall out of equilibrium, they approximately  
decouple from the superfluid dynamics. Next the superfluid 
falls out of equilibrium.
We expect that the  
influence of the quantum critical point on the dynamics (e.g. through the 
Cooper pair size dependence on $\hat\xi$)
will be visible when 
$T\ll T_c(\hat t\,)\sim \sqrt{\tau_0/\tau_Q}T_F$. This ensures that 
there is still a condensate at the time  when 
quantum criticality drives the superfluid  out of equilibrium.
Experimentally one can stop the ramp before the system reaches the estimated classical transition line 
and observe the predicted scaling from a quantum quench.

Since the equilibrium $T_c$ vanishes as $a\rightarrow 0$, 
the system will eventually be swept out of the BCS
superfluid phase.
This raises 
another important question: Is the BCS superfluid-Fermi gas 
classical phase transition still well defined when 
both QPs and the superfluid are out of equilibrium?
We point out that 
crossing a classical critical point may also imprint
 additional  non-equilibrium length scale encoding the 
critical exponents of the classical transition \cite{Zurek}.

In summary,  we have shown that a finite-rate quench allows
for  controlled (through the quench time scale $\tau_Q$) studies
of non-equilibrium dynamics of a BCS superfluid.
The quench imprints non-equilibrium energy  and length scales visible through the 
gap and Cooper pair size.
Our results call for  further development of experimental techniques 
for studies of non-equilibrium cold Fermi gases and 
point out that finite-temperature effects may induce additional scaling behavior of non-equilibrium BCS superfluids 
 in a finite-rate quench.

This work is supported by U.S. Department of Energy through the LANL/LDRD Program.

\bibliographystyle{apsrev4-1}

\begin{thebibliography}{10}
\bibitem{details}%
  \BibitemOpen
  \bibinfo {note} {We parametrize $u_k(t)$ and $v_k(t)$ as $\sin\theta_k
  \exp(i\beta_k/2)$ and $\cos\theta_k \exp(-i\beta_k/2)$ ensuring
  $|u_k|^2+|v_k|^2=1$. The overall phase of the wave-function (\ref{BCSwave})
  drops out from the problem. Dynamical equations for $\theta_k$ and $\beta_k$
  follow from minimization of the quantum action. In our simulations we
  introduce $d_k = \sin2\theta_k\exp(i\beta_k)$ and $g_k = \cos2\theta_k$ and
  Eq.~(\ref{equations}) follows.}%
  \bibAnnoteFile{Stop}{details}%
\bibitem{regularization}%
  \BibitemOpen
  \bibinfo {note} {A regularized expression in real space yields finite results
  equivalent to that from the formula in momentum space.}%
  \bibAnnoteFile{Stop}{regularization}%
\end{thebibliography}
%

\end{document}